\documentclass[12pt]{article}
\usepackage{amsmath}
\usepackage{amssymb}
\usepackage[dvips]{graphicx}
\usepackage[dvips]{epsfig}

 \advance \textwidth by 0.1cm
 \advance \textheight by 0.06cm
\def\##1{{\bf #1}}
\def\=#1{\underline{\underline #1}}
\def\4#1{\underline{\underline{\underline{\underline #1}}}}

\def\.{\mbox{ \tiny{$^\bullet$} }}

\def\le{\left(}
\def\ri{\right)}
\def\les{\left[}
\def\ris{\right]}

\def\c#1{\cite{#1}}

\def\r#1{(\ref{#1})}

\def\epso{\epsilon_{\scriptscriptstyle 0}}
\def\lambdao{\lambda_{\scriptscriptstyle 0}}
\def\muo{\mu_{\scriptscriptstyle 0}}
\def\ko{k_{\scriptscriptstyle 0}}
\def\etao{\eta_{\scriptscriptstyle 0}}

\def\eps{\epsilon}
\def\epsa{\epsilon_a}
\def\epsb{\epsilon_b}
\def\epsc{\epsilon_c}

\def\epsmet{\eps_{met}}

\def\sp{\mathbf s}
\def\pinc{{\mathbf p}_+}
\def\pref{{\mathbf p}_-}

\def\Einc{{\mathbf E}_{inc}({\bf r})}
\def\Erefl{{\mathbf H}_{ref}({\bf r})}
\def\Etr{{\mathbf H}_{tr}({\bf r})}

\def\Hinc{{\mathbf H}_{inc}({\bf r})}
\def\Hrefl{{\mathbf H}_{ref}({\bf r})}
\def\Htr{{\mathbf H}_{tr}({\bf r})}

\def\cpsi{\cos\psi}
\def\spsi{\sin\psi}
\def\ctheta{\cos\theta}
\def\stheta{\sin\theta}

\def\nr{n_\ell}

\def\as{a_s}
\def\ap{a_p}
\def\rs{r_s}
\def\rp{r_p}
\def\ts{t_s}
\def\tp{t_p}

\def\rss{r_{ss}}
\def\rsp{r_{sp}}
\def\rps{r_{ps}}
\def\rpp{r_{pp}}
\def\tss{t_{ss}}
\def\tsp{t_{sp}}
\def\tps{t_{ps}}
\def\tpp{t_{pp}}

\def\Rss{R_{ss}}
\def\Rsp{R_{sp}}
\def\Rps{R_{ps}}
\def\Rpp{R_{pp}}
\def\Tss{T_{ss}}
\def\Tsp{T_{sp}}
\def\Tps{T_{ps}}
\def\Tpp{T_{pp}}

\def\ux{\hat{\mathbf{u}}_x}
\def\uy{\hat{\mathbf{u}}_y}
\def\uz{\hat{\mathbf{u}}_z}

\begin{document}

\begin{center}

{\bf Surface--plasmon wave at the planar interface of a metal film and a
structurally chiral medium}\\

\medskip

{\em Akhlesh Lakhtakia}\\

\medskip

{CATMAS---Computational \& Theoretical
Materials Science Group \\
Department of Engineering Science and
Mechanics \\
Pennsylvania State University, University Park, PA
16802--6812, USA
}

\end{center}

\noindent {\bf Abstract.} The solution of a boundary--value problem formulated for
a modified Kretschmann configuration shows that a surface--plasmon wave can be
excited at the planar interface of a sufficiently thin metal film and a nondissipative structurally chiral medium,
provided the exciting plane wave is $p$--polarized. An estimate of the wavenumber
of the surface--plasmon wave also emerges thereby.


\section{Introduction}

The propagation of electromagnetic waves  localized to the planar interfaces of bulk metals and bulk dielectric materials can be traced back to a hundred years ago \cite{Zenneck}.
Called surface--plasmon waves, they attenuate normally away from the interface,
and are excited only with evanescent waves \cite{SMW75}. 

In the Kretschmann configuration,
the bulk metal is in the form of a thin film of uniform thickness, bounded on one side by a 
high--refractive--index
dielectric material and on the other side by a low--refractive--index dielectric material. A plane wave is
launched in the optically denser dielectric material towards the metal film, in order to excite a surface--plasmon wave at the interface of the metal with the optically rarer dielectric material
\cite{KR1968}. The plane wave must be $p$--polarized, The telltale sign 
is a sharp peak in absorbance (i.e., a sharp trough in  reflectance
without a compensatory peak in transmittance) as the angle of incidence (with respect
to the thickness direction)
of the launched plane wave is changed
\cite{Turb59}.
Because the angle of incidence
for exciting the surface--plasmon wave
is a delicate function of the constitutive properties of all three materials, 
surface--plasmon waves in the  visible and the near--infrared regimes
are exploited for sensing, imaging, and other applications \cite{HYG99,BNC00}.

Generally, the
optically rarer medium is homogeneous, normal to its planar interface
with the metal film at least within the range of the
surface--plasmon field. In this communication, this medium is taken to be
continuously
nonhomogeneous in the thickness direction. Specifically, the optically
rarer medium is structurally chiral, with the axis of
helicoidal nonhomogeneity oriented parallel to the thickness
direction. To my knowledge, the solution of the associated
boundary--value problem has never been reported before, and its application
should draw both
 chiral liquid
crystals \cite[Chap. 4]{Chan} and chiral sculptured thin 
films \cite[Chap. 9]{STFbook} into the plasmonics arena.

The plan of this communication is as follows: Section \ref{theory}
contains a description of a modified Kretschmann configuration, with the optically rarer
medium replaced by a structurally chiral material (SCM) slab of sufficient thickness. The 
combination of the metal
film and the SCM slab is sandwiched between two half--spaces occupied
by the same isotropic dielectric material that is optically denser than the chosen SCM.
A brief description of the electromagnetic boundary--value problem is also
presented. Section~\ref{nrd} contains numerical results to
show that a surface--plasmon wave can be
excited at the planar interface of a metal film and a structurally chiral medium,
provided the incident plane wave is $p$--polarized.

In the following, an $\exp(-i\omega t)$ time--dependence is implicit, with $\omega$
denoting the angular frequency. The free--space wavenumber, the 
free--space wavelength, and the intrinsic impedance of free space are denoted by $\ko=\omega\sqrt{\epso\muo}$,
$\lambdao=2\pi/\ko$, and
$\etao=\sqrt{\muo/\epso}$, respectively, with $\muo$ and $\epso$ being  the permeability and permittivity of
free space. Vectors are in boldface, dyadics underlined twice;
column vectors are in boldface and enclosed within square brackets, while
matrixes are underlined twice and similarly bracketed. Cartesian unit vectors are
identified as $\ux$, $\uy$ and $\uz$.

\section{Theory}\label{theory}
In conformance with the Kretschmann configuration for launching surface--plasmon waves,
the half--space $z\leq 0$ is occupied by a homogeneous, isotropic, dielectric
material described by the relative permittivity scalar $\eps_{\ell}$. Dissipation in this material
is considered to be negligible and its refractive index $n_\ell=\sqrt{\eps_\ell}$
is real--valued and positive. The laminar region 
 $0 \leq z\leq L_{met}$ is occupied by a metal with relative permittivity
 scalar $\epsmet$. A structurally chiral material occupies the region
 $L_{met} \leq z \leq L_{met}+L_{scm}$, the dielectric properties of this
 material being described in the following subsection. Finally, without significant 
 loss of generality in the present context, the half--space
 $z\geq L_{met}+L_{scm}$ is taken to be occupied by the same
 material as fills the half--space $z\leq 0$. All constitutive properties generally depend
 on the angular frequency $\omega$.

A plane
wave, propagating in the half--space
$z \leq 0$ at an angle $\theta\in[0,\pi)$ to the $z$ axis and at an angle $\psi\in[0,2\pi)$
to the $x$ axis in the $xy$ plane, is incident on the metal--coated SCM slab.
The electromagnetic field phasors associated
with the incident plane wave are represented as
\begin{equation}
\left.\begin{array}{l}
\Einc= (\as\,\sp +\ap \,\pinc) \, e^{ i\kappa \le x\cpsi + y\spsi \ri
} \,e^{i\ko\nr z\ctheta}
\\[5pt]
\Hinc=\frac{\nr}{\etao}\, (\as\,\pinc -\ap\, \sp) \, e^{ i\kappa \le x\cpsi + y\spsi \ri
}\, e^{i\ko\nr z\ctheta} 
\end{array}\right\}
\, , \qquad z \leq 0
\, .
\end{equation}
The amplitudes of the $s$-- and the
$p$--polarized components of the incident plane wave, denoted by 
$\as$ and
$\ap$, respectively, are assumed given, whereas 
\begin{equation}
\left.\begin{array}{l}
\kappa =
\ko\nr\stheta\\[5pt]
\sp=-\ux\spsi + \uy \cpsi
\\[5pt]
{\mathbf p}_\pm=\mp\le \ux \cpsi + \uy \spsi \ri \ctheta  + \uz \stheta
\end{array}\right\}
\, .
\end{equation}

The reflected electromagnetic field phasors are expressed as
\begin{equation}
\left.\begin{array}{l}
\Erefl= (\rs\,\sp +\rp \,\pref) \, e^{ i\kappa \le x\cpsi + y\spsi \ri
} \,e^{-i\ko\nr z\ctheta}
\\[5pt]
\Hrefl=\frac{\nr}{\etao}\, (\rs\,\pref -\rp\, \sp) \, e^{ i\kappa \le x\cpsi + y\spsi \ri
} \,e^{-i\ko\nr z\ctheta} 
\end{array}\right\}
\, , \qquad z \leq 0
\, ,
\end{equation}
and the transmitted electromagnetic field phasors  as
\begin{equation}
\left.\begin{array}{l}
\Etr= (\ts\,\sp +\tp\, \pinc) \, e^{ i\kappa \le x\cpsi + y\spsi \ri
} \,e^{i\ko\nr (z-L_\Sigma)\ctheta}
\\[5pt]
\Htr= \frac{\nr}{\etao}\,(\ts\,\pinc -\tp\, \sp) \, e^{ i\kappa \le x\cpsi + y\spsi \ri
}\, e^{i\ko\nr (z-L_\Sigma)\ctheta} 
\end{array}\right\}
\, , \qquad z \geq L_\Sigma
\, ,
\end{equation}
where $L_\Sigma=L_{met}+L_{scm}$.
The reflection amplitudes $\rs$ and $\rp$, as well as the transmission
amplitudes $\ts$ and $\tp$, have to be determined by the solution of
a boundary--value problem.

\subsection{Constitutive Relations of the SCM}
The frequency--domain electromagnetic constitutive relations of the SCM
slab can be written as \cite{STFbook}
\begin{equation}
\left.\begin{array}{l}
\#D(\#r) = \epso\,\=\epsilon_{scm}(z)\.\#E(\#r)
\\[4pt]
\#B(\#r) = \muo\#H(\#r) 
\end{array} \right\}\, , \quad L_{met}\leq z \leq L_\Sigma\,.
\end{equation}
The frequency--dependent relative permittivity dyadic $\=\epsilon_{scm}(z)$  is
factorable as
\begin{equation}
\nonumber
\=\epsilon_{scm}(z) =  \=S_z(z-L_{met})\.\=S_y(\chi)\.\=\epsilon^{
ref}_{scm}
\.\=S_y^T(\chi)\.\=S_z^T(z-L_{met})
\, , \quad L_{met}\leq z \leq L_\Sigma\,,
\end{equation}
where  the reference relative permittivity dyadic
\begin{equation}
\=\epsilon_{scm}^{ref}= \epsa  \, \uz\uz  + \epsb \, \ux\ux
+ \epsc \, \uy\uy\,.
\end{equation}
The  
dyadic function
\begin{equation}
\=S_z(z)=\le \ux\ux + \uy\uy \ri \cos{\le \frac{\pi z}{\Omega} \ri}\\
+h\le \uy\ux -
\ux\uy \ri \sin{\le \frac{\pi z}{\Omega} \ri}+\uz\uz \, ,
\end{equation}
contains $2\Omega$ as the structural
period and $h=\pm 1$ as the structural--handedness parameter;
thus,  the SCM is
helicoidally nonhomogeneous along the $z$ axis. The tilt dyadic
\begin{equation}
\=S_y(\chi)=\le \ux\ux + \uz\uz \ri \cos{\chi}
+\le \uz\ux -
\ux\uz \ri \sin{\chi}+\uy\uy \, 
\end{equation}
involves the angle   $\chi\in[0,\pi/2]$.
The superscript $^T$ denotes the transpose.

\subsection{Boundary--Value Problem}
The procedure to determine the amplitudes $r_s$, $r_p$,
$t_s$, and $t_p$ in terms of $a_s$ and $a_p$ is standard
\c{STFbook}. It suffices to state here that the following set
of 4 algebraic equations emerges (in matrix notation):
\begin{equation}
\left[\begin{array}{l} t_s\\t_p\\ 0 \\0\end{array}\right]=
[\=K]^{-1}\cdot[\=B_{scm}]\cdot[\=M^\prime_{scm}]\cdot
\exp\left({i[\=P_{met}]L_{met}}\right)\cdot[\=K]\cdot
\left[\begin{array}{l} a_s\\a_p\\ r_s \\r_p\end{array}\right]\,.
\label{finaleq}
\end{equation}
The procedure to compute the 4$\times$4 matrix
$[\=M^\prime_{scm}]$ is far too cumbersome for reproduction
here, the interested reader being referred to \cite[Sec. 9.2.2]{STFbook}.
The 4$\times$4 matrix $[\=K]$ depends on the refractive index $\nr$ as well as
the angles $\theta$ and $\psi$ as follows:
\begin{equation}
[\=K] =
\left[ \begin{array}{cccc}
-\sin\psi & -\cos\psi\cos\theta  & -\sin\psi & \cos\psi\cos\theta \\
\cos\psi & -\sin\psi\cos\theta  & \cos\psi & \sin\psi\cos\theta \\ 
-\left(\frac{\nr}{\etao}\right)\cos\psi\cos\theta & \left(\frac{\nr}{\etao}\right)\sin\psi  
&\left(\frac{\nr}{\etao}\right)\cos\psi\cos\theta &
\left(\frac{\nr}{\etao}\right)\sin\psi \\  
-\left(\frac{\nr}{\etao}\right)\sin\psi\cos\theta &
-\left(\frac{\nr}{\etao}\right)\cos\psi &\left(\frac{\nr}{\etao}\right)\sin\psi\cos\theta & -\left(\frac{\nr}{\etao}\right)\cos\psi
\end{array}\right]\,.
\end{equation}
The remaining two matrixes appearing in \r{finaleq} are
\begin{equation}
[\=B_{scm}]=
\les\begin{array}{cccc}
\cos{\left(\pi L_{scm}/\Omega \right)} & -h\,\sin{\left(\pi L_{scm}/\Omega \right)} & 0 & 0  \\
h\,\sin{\left(\pi L_{scm}/\Omega \right)} & \cos{\left(\pi L_{scm}/\Omega \right)} & 0 & 0  \\
0 & 0 & \cos{\left(\pi L_{scm}/\Omega \right)} & -h\,\sin{\left(\pi L_{scm}/\Omega \right)} \\  
0 & 0 & h\,\sin{\left(\pi L_{scm}/\Omega \right)} & \cos{\left(\pi L_{scm}/\Omega \right)} \\  
\end{array}\ris\,
\end{equation}
and
\begin{eqnarray}
\nonumber
[\=P_{met}]&=&\les\begin{array}{cccc}
0 & 0 & 0 & \omega\muo \\[4pt]
0 & 0 & -\omega\muo & 0 \\[4pt]
0 & -\omega\epso\epsmet & 0 &0\\[4pt]
\omega\epso\epsmet &0 & 0 & 0
\end{array}\ris\\[5pt]
\nonumber &&
+\,\frac{\kappa^2}{\omega\epso\epsmet}\,
\les\begin{array}{cccc}
0 & 0 & \cpsi\spsi & -\cos^2\psi  
\\[4pt]
0 & 0 & \sin^2\psi  &- \cpsi\spsi \\[4pt]
0 & 0& 0 & 0\\[4pt]
0 & 0 & 0 & 0
\end{array}\ris
\\[5pt]
\label{Pmetal}
&&\qquad
+\,\frac{\kappa^2}{\omega\muo}\,
\les\begin{array}{cccc}
0 & 0 & 0 & 0\\[4pt]
0 & 0 & 0 & 0\\[4pt]
-\cpsi\spsi & \cos^2\psi  & 0 
& 0\\[4pt]
-\sin^2\psi & \cpsi\spsi  & 0 & 0
\end{array}\ris\,.
\end{eqnarray}

The solution of \r{finaleq} yields
the reflection and transmission coefficients that appear as the elements of the 
2$\times$2 matrixes in the following
relations:
\begin{equation}
\label{eq9.55}
\les\begin{array}{c}\rs\\\rp\end{array}\ris
=
\les\begin{array}{cc}\rss & \rsp\\\rps & \rpp\end{array}\ris
\,
\les\begin{array}{c}\as\\\ap\end{array}\ris\,,
\qquad
\les\begin{array}{c}\ts\\\tp\end{array}\ris
=
\les\begin{array}{cc}\tss & \tsp\\\tps & \tpp\end{array}\ris
\,
\les\begin{array}{c}\as\\\ap\end{array}\ris\,.
\end{equation}
Co--polarized coefficients have both subscripts identical, but
cross--polarized coefficients do not. The square of the magnitude
of a reflection or transmission coefficient is the corresponding
reflectance or transmittance;  thus, $\Rsp = \vert\rsp\vert^2$ is
the reflectance corresponding to the reflection coefficient $\rsp$,
and so on.
The principle of conservation of energy mandates
the constraints
$
\Rss + \Rps + \Tss + \Tps \leq 1$ and
$\Rpp + \Rsp + \Tpp + \Tsp \leq 1
$,
the inequalities turning to equalities only in the
absence of dissipation in the region $0<z<L_\Sigma$. 

\section{Numerical Results and Discussion}\label{nrd}
All eight  reflectances and transmittances at the free--space
wavelength $\lambdao=633$~nm were calculated
as functions of the angles $\theta$ and $\psi$. The SCM
was chosen to possess the following parameters: $\epsa=2.7$,
$\epsb=3.0$, $\epsc=2.72$, $\chi=30^\circ$, $\Omega=200$~nm,
and $h=\pm1$. The relative permittivity of the ambient medium was
chosen to be $\eps_\ell=5$, and that of the metal (typ. aluminum)
as $\epsmet=-56+i21$. For the chosen constitutive parameters, the constraints $
\Rss + \Rps + \Tss + \Tps = 1$ and
$\Rpp + \Rsp + \Tpp + \Tsp = 1
$
hold in the absence of the metal film.

\begin{figure}[!ht]
\centering \psfull
\epsfig{file=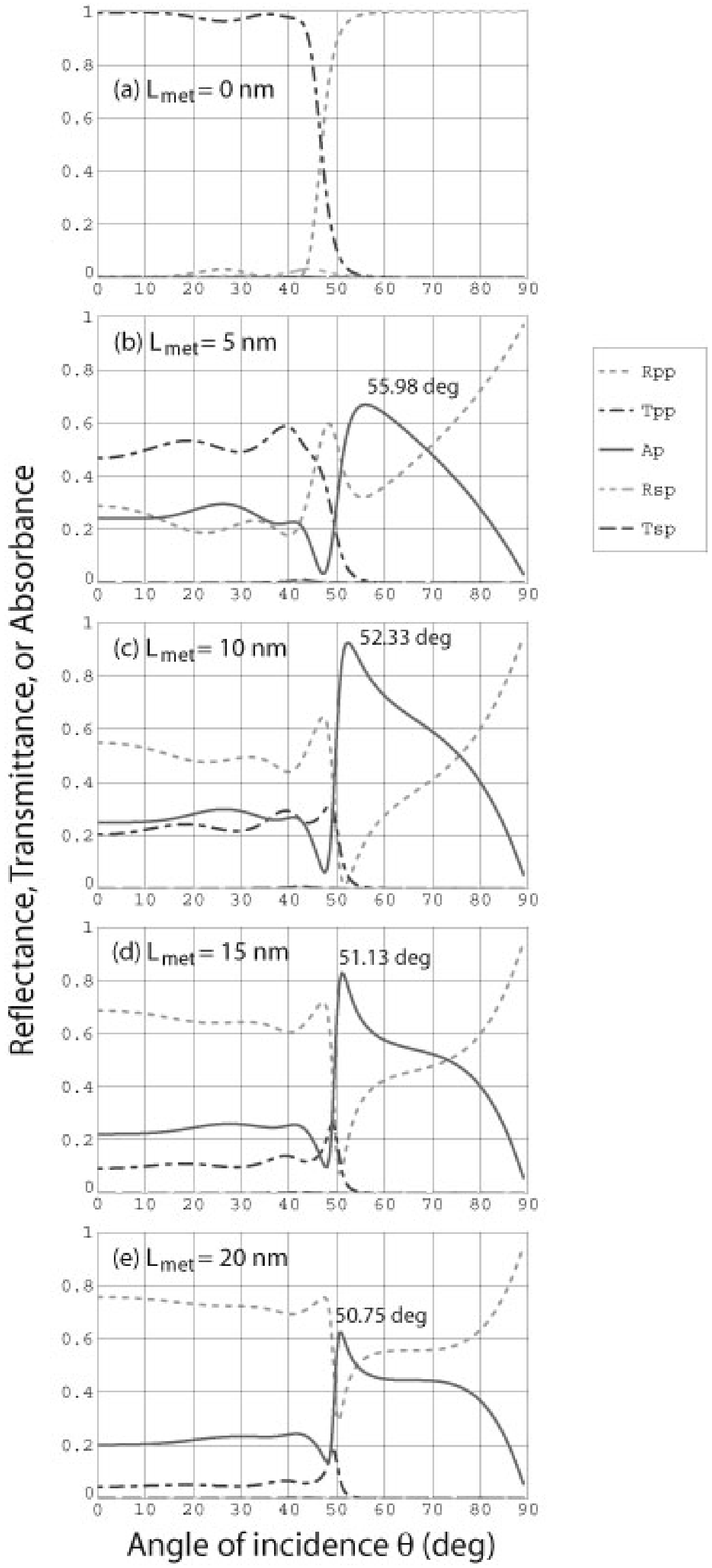,
height=16cm
}
\caption{Reflectances ($\Rpp$ and $\Rsp$), transmittances
($\Tpp$ and $\Tsp$), and the absorbance  as functions of $\theta$ when $\psi=0^\circ$,
$\lambdao=633$~nm,
and the incident plane wave is $p$--polarized.
The SCM is described by the following parameters:
$\epsa=2.7$,
$\epsb=3.0$, $\epsc=2.72$, $\chi=30^\circ$, $\Omega=200$~nm, $h=\pm1$,
and $L_{scm}=2\Omega$. The relative permittivity of the metal is
$\epsmet=-56+i21$,
and that of the ambient medium is
$\eps_\ell=5$.  (a) $L_{met}=0$, (b) $L_{met}=5$~nm,
(c) $L_{met}=10$~nm, (d) $L_{met}=15$~nm, and (e)  $L_{met}=20$~nm.
The values of $\theta$ for maximum $A_p$ are identified for different non--zero
values
of $L_{met}$ in the plots. 
}
\label{figp1period}
\end{figure}

Figure~\ref{figp1period} shows the variations of the reflectances
 ($\Rpp$ and $\Rsp$), transmittances
($\Tpp$ and $\Tsp$), and the absorbance 
\begin{equation}
A_p=1-
(\Rpp + \Rsp + \Tpp + \Tsp)
\end{equation}
 with $\theta$ when $\psi=0^\circ$,
and the incident plane wave is $p$--polarized. The SCM is 1-period
thick (i.e., $L_{scm}=2\Omega$), whereas the thickness of the metal
film varies from 0 to 20~nm in steps of 5~nm. 
A rapid increase
in the absorbance $A_p$ indicates the excitation of a surface--plasmon
wave \cite{Masud}. The values of $\theta$ for maximum $A_p$ are identified for different non--zero
values
of $L_{met}$ in Fig.~\ref{figp1period}. For instance,
the absorbance equals $0.93$
at $\theta=52.33^\circ$, when $L_{met}=10$~nm. As $L_{met}$ increases,
the maximum--absorbance value of $\theta$ decreases slightly,
whereas the maximum absorbance decreases as well.

\begin{figure}[!ht]
\centering \psfull
\epsfig{file=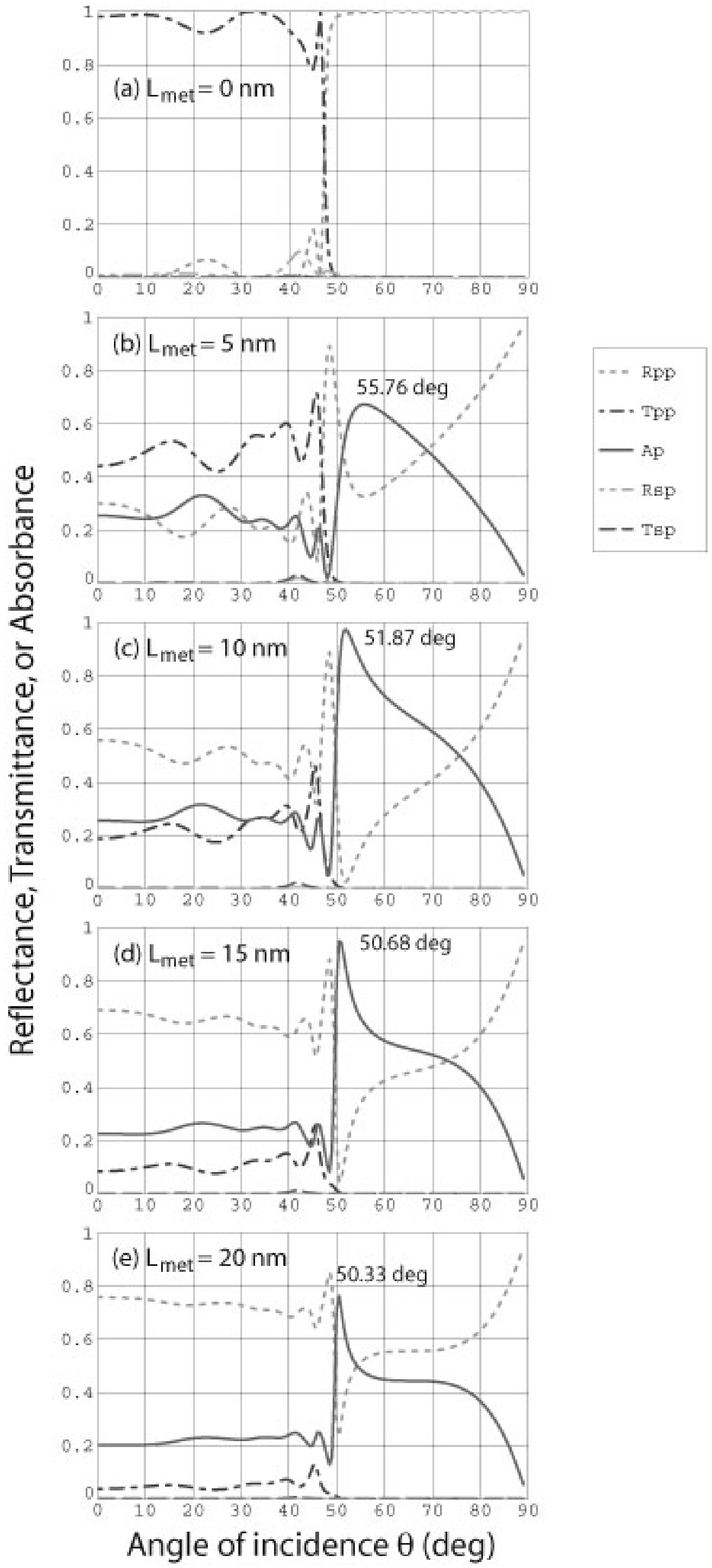,
height=16cm
}
\caption{Same as Fig.~\ref{figp1period}, except that $L_{scm}=4\Omega$. }
\label{figp2periods}
\end{figure}

\begin{figure}[!ht]
\centering \psfull
\epsfig{file=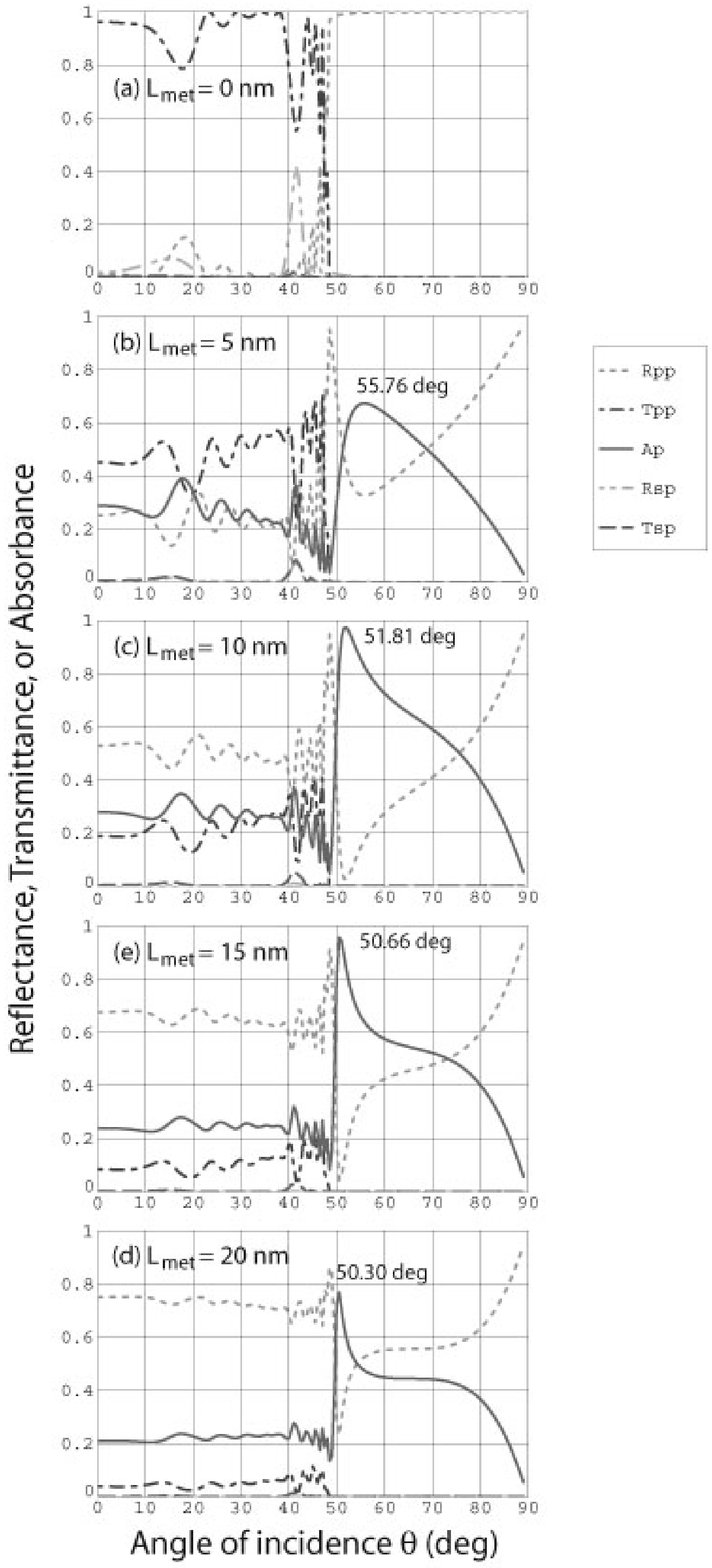,
height=16cm
}
\caption{Same as Fig.~\ref{figp1period}, except that $L_{scm}=10\Omega$. }
\label{figp5periods}
\end{figure}

The calculations for Fig.~\ref{figp1period} were repeated for
higher values of $L_{scm}/\Omega$. As the thickness
of the SCM slab was increased, the maximum--$A_p$ value
of $\theta$ for a specific value of $L_{met}$ began to
converge. This is exemplified by the plots of $A_p$ vs.
$\theta$ in Fig.~\ref{figp2periods} for  $L_{scm}/\Omega =4$
and  Fig.~\ref{figp5periods} for  $L_{scm}/\Omega=10$. 
Thus, the maximum value of $A_p$ is $0.975$ 
at $\theta=51.87^\circ$ in Fig.~\ref{figp2periods}(b)
and also  at $\theta=51.81^\circ$ in Fig.~\ref{figp5periods}(b),
both for $L_{met}=10$~nm.  

A comparison of the three figures
indicates that the 5--period thick SCM slab is sufficiently thick as
to be equivalent to a SCM half--space, which would be required
in the usual theoretical treatment of the (unmodified) Kretschmann configuration
\cite{SMW75}. Parenthetically, the planewave response of
a SCM half--space cannot be obtained unless the wavevector of the incident
plane wave is aligned parallel to the $z$ axis \cite{LM02}, because
a sufficiently general eigenmodal decomposition of the electromagnetic fields is unavailable \cite[Chap. 9]{STFbook}.

Were the SCM 
replaced by an isotropic dielectric material of relative
permittivity $\eps_{iso}$ and the metal
film were absent, total internal reflection would occur for 
$\theta\geq\sin^{-1}\sqrt{\eps_{iso}/\eps_\ell}$. Then, $\sin^{-1}\sqrt{\eps_{iso}/\eps_\ell}$ is the critical angle and $\ko\sqrt{\eps_{iso}}$ is an estimate of the
wavenumber of the surface--plasmon wave \cite{Masud}. But
a simple formulation of
a ``critical angle" is not possible with the SCM. A useful estimate can, however,
be made, by setting $\eps_{iso}= {\rm max}\left(\epsa,\epsb,\epsc\right)$,
whereby the ``critical angle" equals $\sin^{-1}\sqrt{\epsc/\eps_\ell}=50.77^\circ$ for the chosen parameters. Figures~\ref{figp2periods}(b)--(e) and \ref{figp5periods}(b)--(e)
indicate that the surface--plasmon wave is indeed excited in the neighborhood
of this estimate of this ``critical angle", which is also ratified by the plots for $L_{met}=0$
in
Figs.~\ref{figp2periods}(a) and \ref{figp5periods}(a).

\begin{figure}[!ht]
\centering \psfull
\epsfig{file=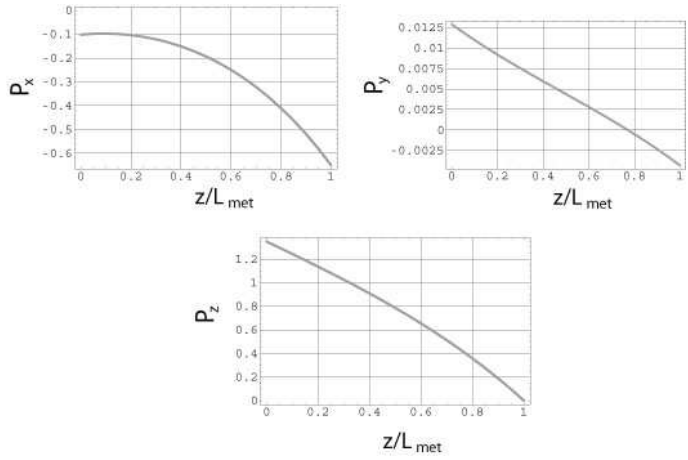,
height=6cm
}
\caption{Cartesian components of the time--averaged Poynting vector $\#P(z)$
in the metal film vs. $z\in(0,L_{met})$ when a surface--plasmon wave
has been excited. The conditions are the same as for Fig.~\ref{figp5periods}(c),
except that $h=1$. For $h=-1$,  the $y$--directed component of $\#P(z)$
is different in sign but not in magnitude.
}
\label{figPoynting}
\end{figure}

In order to confirm the excitation of a surface--plasmon wave
at the interface of the metal and the SCM, the time--averaged
Poynting vector $\#P(z) = (1/2) {\rm Re} \left[\#E(z)\times \#H^\ast(z)\right]$
was plotted against $z\in(0,L_{met})$ for all calculations reported
in the previous figures. Shown in Fig.~\ref{figPoynting} are the
cartesian components of $\#P(z)$ vs. $z$ in the metal film,
when $\theta=51.81^\circ$, $L_{met}=10$~nm, $L_{scm}=10\Omega$,
and all other parameters are the same as for Fig.~\ref{figp5periods}.
The magnitude of $P_z$ decreases and that of $P_x$ increases, both
monotonically, as one traverses the metal film from the interface with the medium
of incidence ($z=0$) to the interface with the SCM ($z=L_{met}$). Clearly thus,
the presence of surface--plasmon
wave localized to the interface $z=L_{met}$ is confirmed. 

Figure~\ref{figPoynting} also shows the effects of the anisotropy and the
structural handedness
of the SCM. These effects are manifested  in the $y$--directed component of $\#P(z)$
in the metal film. Were the SCM to be replaced by an isotropic material,
this component of $\#P(z)$ would be identically zero for all $z$. Also,
the sign of this component depends on whether $h=1$ or $h-1$.

Although all numerical results presented were calculated for $\psi=0^\circ$,
calculations were made for other values of $\psi$ as well.
No significant effect of $\psi$ on the maximum--$A_p$ value of $\theta$
was detected.

\begin{figure}[!ht]
\centering \psfull
\epsfig{file=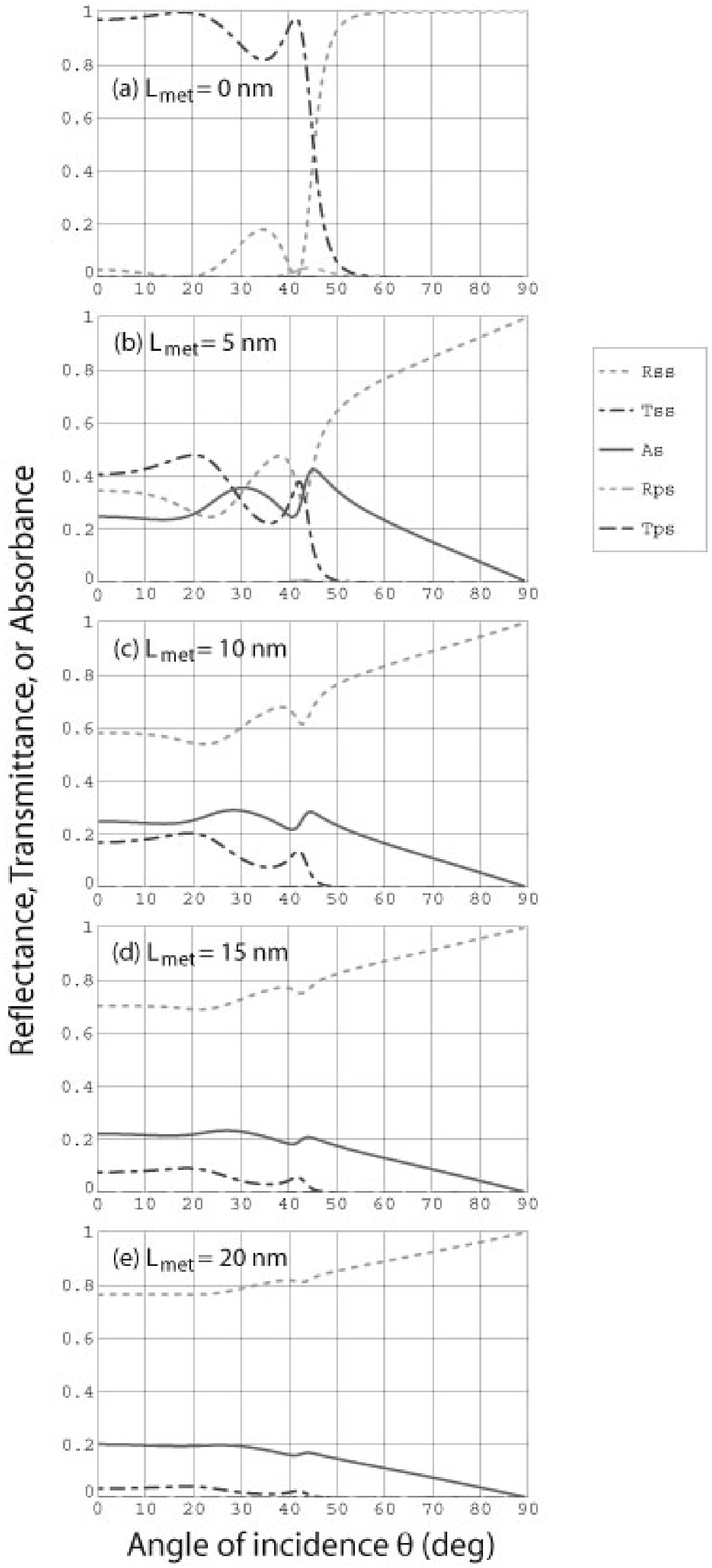,
height=16cm
}
\caption{Reflectances ($\Rss$ and $\Rps$), transmittances
($\Tss$ and $\Tps$), and the absorbance  as functions of $\theta$ when $\psi=0^\circ$,
$\lambdao=633$~nm,
and the incident plane wave is $s$--polarized.
The SCM is described by the following parameters:
$\epsa=2.7$,
$\epsb=3.0$, $\epsc=2.72$, $\chi=30^\circ$, $\Omega=200$~nm, $h=\pm1$,
and $L_{scm}=2\Omega$. 
The relative permittivity of the metal is
$\epsmet=-56+i21$,
and that of the ambient medium is
$\eps_\ell=5$. 
(a) $L_{met}=0$, (b) $L_{met}=5$~nm,
(c) $L_{met}=10$~nm, (d) $L_{met}=15$~nm, and (e)  $L_{met}=20$~nm.}
\label{figs1period}
\end{figure}

Figure~\ref{figs1period} shows the variations of the relevant  reflectances
and transmittances, and of the absorbance 
\begin{equation}
A_s=1-
(\Rss + \Rps + \Tss + \Tps)\,,
\end{equation}
with $\theta$ for the same parameters as for Fig.~\ref{figp1period},
except that the incident plane wave is $s$--polarized. Evidence of the excitation of
a surface--plasmon wave is absent from this figure, just as it would
be if the SCM slab were to be replaced by a slab made of an isotropic
dielectric material \cite{Masud}. Calculations for higher values of the ratio 
$L_{scm}/\Omega$
also did not reveal the existence of a surface--plasmon wave for $s$--polarized
incidence.

To conclude, the solution of a boundary--value problem formulated for
a modified Kretschmann configuration shows that a surface--plasmon wave can be
excited at the planar interface of a 
sufficiently thin metal film and a nondissipative  structurally chiral medium,
provided that (i) the incident plane wave is $p$--polarized, and (ii) the wavenumber (i.e., $\kappa$) of
the surface--plasmon wave roughly equals $\ko\sqrt{{\rm max}(\epsa,\epsb,\epsc)}$.
The estimated wavenumber of the surface--plasmon wave may have to be obtained
graphically (by setting $L_{met}=0$), if $\eps_{a,b,c}$ are very different from each other.

\bigskip 

\noindent {\bf Acknowledgment.} This work was supported in part by the Charles
Godfrey Binder Endowment.


\end{document}